\documentstyle[12pt]{article}
\newcommand{\tr}[1]{\,{\rm tr}\,#1\,}
\newtheorem{lemma}{Lemma}
\newtheorem{theor}{Theorem}
\newtheorem{defin}{Definition}

\begin{document}
\title{
\begin{flushright}
{\small SMI-5-94\\
HEP-TH/9401127 }
\end{flushright}
\vspace{2cm}
Poisson-Lie structures \\ on the external algebra of $SL(2)$ \\ and their
quantization}
\author{I.Ya Aref'eva\thanks{Steklov Mathematical Institute, Vavilov 42, GSP-1,
117966,
Moscow,
Russia;$~~~~~~~~~~~~$ $~~~~~~$arefeva@qft.mian.su},$~~$
G.E.Arutyunov \thanks{Steklov Mathematical Institute, Vavilov 42, GSP-1,
117966, Moscow,
Russia; arut@qft.mian.su}
 \\
and\\
P.B.Medvedev \thanks
{Institute of Theoretical and Experimental Physics,
117259 Moscow, Russia}
}
\date {January 1994~}

\maketitle
\begin{abstract}
All possible graded Poisson-Lie  structures on the external algebra of $SL(2)$
are described.
We prove that differential Poisson-Lie structures prolonging the Sklyanin
brackets do not exist on $SL(2)$.
There are two and only two graded Poisson-Lie structures
on $SL (2)$ and  neither of them can be obtained by a reduction of
graded Poisson-Lie  structures on the external algebra of $GL(2)$.
Both of them can be quantized and
as a result we get new  graded algebras of quantum right-invariant forms on
$SL_q(2)$ with three generators.
\end{abstract}

\newpage

\section{Introduction}
Recently the problem of constructing differential calculi
on the quantum
{\it special} linear group $SL_q(N)$ \cite{Fad,Dr} has attracted
considerable attention \cite{Wor1}-\cite{IS}. A peculiar attention was
paid to constructing bicovariant differential calculi on these quantum
groups \cite{Wor}-\cite{IS}. Just the requirement of bicovariance, i.e.
invariance under quantum version of right and left group transformations,
aside from the
obvious geometrical meaning has a physical interpretation
\cite{AV}-\cite{ArA}.
As it was recognized the proposed solution reveals some strange
properties. Namely, the bicovariant differential complex involves
an extra element which has no natural classical counterpart in $SL(N)$.
In this paper we construct a quantum algebra of bicovariant forms associated
with
$SL_q(2)$ that is free from this defect.

Let us note that one can say that there are two main approaches to
differential calculus on noncommutative spaces.
The first is the Connes approach \cite{KON} in which
a complex of differential forms is constructed and the operator
$\bf d$ of exterior derivative plays a fundamental role. Such an approach
was used for quantum groups by Woronowicz \cite{Wor1}.

Another approach can be formulated following the  Faddeev idea
that all the objects in the theory of quantum groups
should appear naturally as the result of quantization
of appropriate Poisson
brackets in the theory of Lie groups \cite{FF}.
This point of
view pushes us to start a search for an algebra of quantum forms
with finding natural Poisson-Lie structures on the external
algebra of ordinary differential forms. These Poisson-Lie
structures being quantized should give as a result
algebras of quantum differential forms.
The ${\bf d}$ operator does not play a fundamental role in this approach,
moreover, generally there are no reasons to be sure that quantum
$\bf d$ exists.

Just this point of view was advocated in the paper \cite{AM}.
It was found that
the external algebra on $GL(N)$ can be equipped with the graded
Poisson-Lie structures and just their quantization produces
the bicovariant differential calculi on $GL_q(N)$ \cite{AM}.
This forces us to search directly for
graded Poisson-Lie structures on $SL(N)$ whose subsequent quantization
would give the bicovariant calculi on $SL_q(N)$ rather than to discuss
possible reductions of $GL_q(N)$-calculi to the $SL_q(N)$ case.

We are going to describe all possible graded Poisson-Lie  structures on
the external algebra of $SL(2)$.
We will prove that {\em differential} Poisson-Lie structures prolonging the
Sklyanin brackets do not exist on $SL(2)$. This key observation explains
why does not exist {\it bicovariant } differential calculi on $SL_q (2)$
which is in one to one correspondence with their classical counterpart.
The absence of a unique preferred
Poisson structure
(differential Poisson-Lie structure) provokes the
discussion of different
Poisson brackets on the external algebra
by relaxing some requirements on Poisson
structures and their further quantization. The more natural
possibility which we are going to deal with
in this paper consists in
consideration of a graded Poisson-Lie structure without requiring this
structure to be the differential one.

We will find that there are two and only two graded Poisson-Lie structures
on $SL (2)$. Neither of them can be obtained by a reduction of
graded Poisson-Lie  structures on the external algebra of $GL(2)$.
Then we prove that both of them can be quantized.
As a result we get two graded algebras of quantum right-invariant forms on
$SL_q(2)$. As we could have expected from classical treatment
these algebras of quantum forms on
$SL_q(2)$ do not admit the operator $\bf d$ with standard properties.

The paper is organized as follows. In section 2 we fix our notations
and introduce the notion of a graded Poisson-Lie algebra.
The rest of the section is devoted to the description of all such
structures  on the external algebra of $SL(2)$.
In section 3 the quantization of these  graded Poisson-Lie
structures is presented.

\section{Graded Poisson-Lie structure associated with $SL(2)$}
\subsection{Definitions}
The function algebra ${\cal A}$ on
$SL(2,C)$ is defined as the free unital associative algebra generated by
the entries of the
matrix:
\begin{equation}
T=||t_{m}^{n}||=
\left(\begin{array}{ll}a & b\\ c & d\end{array}\right),
\label{ttt}
\end{equation}
modulo the additional relation  $\det{T}=ad-bc=1.$

Recall, that $\cal A$ has
the natural
Hopf algebra structure
with comultiplication $\Delta$, the counit $\epsilon$
and the antipode $S$:
$$\Delta(T) =T\otimes T,~~\epsilon(T)=I,~~S(T)=T^{-1}.$$
Specifying $T$ as the matrix (\ref{ttt}) we have for the comultiplication
$\Delta$:
\begin{equation}
\Delta \left(\begin{array}{ll}a & b\\ c &
d\end{array}\right)= \left(\begin{array}{ll}a & b\\ c &
d\end{array}\right)\otimes \left(\begin{array}{ll}a & b\\ c &
d\end{array}\right).
\label{uu}
\end{equation}

According to the general theory of Poisson-Lie groups \cite{Dr,Fad,Skl,Sem}
the function algebra on $SL(2)$ can be supplied with the Poisson
structure $\{,\}$ compatible with the comultiplication $\Delta$, {\em i.e.}
satisfying the condition
\begin {equation}
\label {csk}
\Delta\{f,h\}=\{\Delta f, \Delta h\},~~f,h\in \cal A.
\end   {equation}
In terms of generators $t_{i}^{j}$ this Poisson structure  is given by
the Sklyanin bracket \cite{Skl,Skl1}:
\begin{equation}
\begin{array}{lll}
\{a,b\}=-ab, & \{b,d\}=-bd, &  \{c,b\}=0, \\
\{a,c\}=-ac, & \{c,d\}=-cd, &  \{a,d\}=-2bc.
\end{array}
\label{sk}
\end{equation}

Let
\begin{equation}
\theta=\left(\begin{array}{rr}\theta^0 &\theta^1\\ \theta^2
& -\theta^0\end{array}\right)
\label{theta}
\end{equation}
be the canonical right-invariant
one-form on $SL(2)$ taking value in $sl(2)$.
The components
$\theta^i,~i=0,1,2$ being the scalar right invariant one-forms can be
viewed as the generators of the external algebra on $SL(2)$.
In what follows we will refer to
$\theta$ as to the Maurer-Cartan form on $SL(2)$. Note, that the choice of
right-invariant forms is a matter of convention
and they may be replaced by left-invariant ones.

Let us define now the basic object of our study. This is a
$Z_2$-graded algebra
$\cal M$ generated by even $t_{m}^{n}\in {\cal A}$ and
odd $\theta^i$ generators modulo the defining relations:
\begin{equation}
t_{m}^{n}t_{k}^{l}=t_{k}^{l}t_{m}^{n},~~
t_{m}^{n}\theta^i=\theta^it_{m}^{n},~~
\theta^i\theta^j=-\theta^j\theta^i,
\label{hh}
\end{equation}
$$\det{T}=ad-bc=1.$$
Let $\Omega$ (the algebra of external forms) is the subalgebra
of $\cal M$ generated by $\theta^i$.
Clearly, $\cal A$ is also the subalgebra of $\cal M$.

Now our task is to supply $\cal M$ with a coalgebra structure,
{\em i.e.} one needs to define a
homomorphism $\Delta
:{\cal M}\rightarrow{\cal M}\otimes{\cal M}$ which satisfies
the axiom of coassociativity. Obviously, we require $\Delta$
when being restricted on the subalgebra $\cal A$ to coincide with the
comultiplication (\ref{uu}).
Since $\Delta$ when acting on $\Omega$ should encode the
transformation law of the right-invariant
form under the left and right group
multiplications we define two different comultiplications
which we call $\Delta_R$ and $\Delta_L$ \footnote{
In the Woronowicz terminology \cite{Wor} $\Omega$ is called
a bicovariant-bimodule over $\cal A$}:
$$
\Delta_R:\Omega  \rightarrow {\cal A}\otimes \Omega,
$$
$$
\Delta_L:\Omega \rightarrow \Omega \otimes {\cal A}.
$$
Indeed, the components $\theta^i$ of the Maurer-Cartan
form transforms under the left and right group transformations:
$g\rightarrow gg_{1}$ and
$g\rightarrow g_1g$, $g,~g_1\in SL(2)$, as
\begin{equation}
R^{*}_{g_{1}}\theta_{gg_1}=\theta_{g}~~~ \mbox{and}~~~
L^{*}_{g_{1}}\theta_{g_1g}=Ad_{g_1}\theta_{g}
\label{rl}
\end{equation}
respectively. Therefore, in accordance with (\ref{rl})
we define

\begin {equation}
\label {rm}
\Delta_R\theta^i=\theta^i\otimes I
\end   {equation}
and
\begin {equation}
\label {lm}
\Delta_L\theta^i=(Ad_ge_j)^{i} \otimes \theta^j ,
\end   {equation}
where $\{e_j\}$ is a
basis in $sl(2)$.

In addition to $\Delta_{L,R}$ one can consider the comultiplication
$\Delta$ defined as the sum:
\begin{equation}
\Delta_G=\Delta_{L}+\Delta_{R}
\label{dd}
\end{equation}
Concerning this $\Delta_G$ one can supply the $\cal M$
by a counit $\epsilon$ defined on $\theta^i$ as
$\epsilon(\theta^i )=0$. Thus, there are three natural ways to prolong
the comultiplication $\Delta$ from $\cal A$ to the hole algebra
$\cal M$: $\Delta_{L}$, $\Delta_{R}$ and $\Delta_G$. Note, that just the
comultiplication (\ref{dd}) was used for $GL(N)$ \cite{AM}.

Using the notations (\ref{ttt}) for the
generators of $\cal A$ equation  (\ref {lm}) can be rewritten as
\begin{equation}
\Delta_L\theta^0=(1+2bc)\otimes\theta^0-ac\otimes \theta^1
+bd\otimes\theta^2,
\label{df}
\end{equation}
\begin{equation}
\Delta_L\theta^1=-2ab\otimes\theta^0+a^2\otimes \theta^1
-b^2\otimes\theta^2,
\label{df1}
\end{equation}
\begin{equation}
\Delta_L\theta^2=2cd\otimes\theta^0-c^2\otimes \theta^1
+d^2\otimes\theta^2.
\label{df2}
\end{equation}

Let us recall that the graded algebra $\cal M$ becomes a
graded Poisson algebra if we define
a bilinear operation $\{,\}:$ $~{\cal M}\otimes {\cal M}$
 $\to {\cal M}$ which satisfies

i) the super Jacobi identity:
\begin{equation}
(-1)^{\deg{x}\deg{z}}\{\{ x,y\}, z \}+(-1)^{\deg{y}\deg{z}} \{\{ z,x\}, y
\}+ (-1)^{\deg{x}\deg{y}}\{\{ y,z\}, x \}=0,
\label{ya}
\end{equation}

ii) the graded Leibniz rule:
\begin{equation}
\{x\cdot y, z \}=x\{y, z \}+(-1)^{\deg{y}\deg{z}}\{x, z \}y,
\label{ya1}
\end{equation}

iii) the graded symmetry property:
\begin{equation}
\{x, y \}=(-1)^{\deg{x}\deg{y}+1}\{y, x \} ,~~~ \deg{\{x, y
\}}=(\deg{x}+\deg{y})\bmod 2.
\label{ya2}
\end{equation}

The graded Poisson brackets on $\cal M$ can be extended to
the brackets ${\cal M}\otimes{\cal M}$ in a natural way:
\begin{equation}
\{ x\otimes y,z\otimes w\}_{{\cal M}\otimes {\cal M}}=
(-1)^{\deg{y}\deg{z}}\{ x,z\}_{\cal M}\otimes yw+
(-1)^{\deg{y}\deg{z}}xz\otimes \{ y,w\}
\label{zzx}
\end{equation}
for any elements $x,y,z,w\in {\cal M}$.
\begin {defin}
A graded Poisson algebra $\cal M$ is a graded Poisson-Lie algebra if it
 admits a Poisson-Lie structure, i.e. the
following compatibility conditions are satisfied:

1)for any elements $f,h \in \cal A$
\begin{equation}
\Delta\{f,h\}_{\cal M}=\{\Delta f, \Delta h\}_{{\cal M}\otimes{\cal M}},
\label{cova}
\end{equation}

2)the brackets involving the odd generators $\theta^i$ are
covariant with respect to the both actions of $\Delta_L$ and $\Delta_R$, {\em
i.e.}:
\begin{equation}
\Delta_{L,R} \{\theta^i,f\}_{\cal M}=\{\Delta_{L,R}
\theta^i,\Delta f\}_{{\cal M}\otimes{\cal M}},~~i=0,1,2,~~
f\in {\cal A}
\label{*}
\end{equation}
\begin{equation}
\Delta_{L,R} \{\theta^i,\theta^j\}_{\cal M}=\{\Delta_{L,R}
\theta^i,\Delta_{L,R} \theta^j\}_{{\cal M}\otimes{\cal M}},~~i,j=0,1,2.
\label{**}
\end{equation}
\end{defin}

We also say that a graded Poisson-Lie structure is of the first order if
(\ref{cova}) and (\ref {*}) are satisfied, and of the
second order if (\ref{cova}) and (\ref {**}) take place.

Now according to our general strategy
we look for a graded Poisson structure (brackets) on $\cal M$
obeying the natural covariance conditions (\ref {cova}), (\ref{*}) and
(\ref{**}).

One can also search for a graded Poisson
structure covariant under the action of $\Delta_G$ given by (\ref{dd}).
Note that for $GL(N)$ the graded Poisson-Lie structures are covariant with
respect to $\Delta_G$ \cite{AM}, but in
general a Poisson-Lie algebra can appear to be
a $\Delta_G$-noncovariant one.

\subsection{$\{\theta, t\}$ brackets.}
Due to the grading
requirement (\ref{ya2}) we can write down
the general expression for the brackets as
\begin{equation}
\{\theta^i,
a\}=C_{aj}^{i}\theta^j,~~ \{\theta^i, b\}=C_{bj}^{i}\theta^j,~~
\{\theta^i, c\}=C_{cj}^{i}\theta^j,~~
\{\theta^i, d\}=C_{dj}^{i}\theta^j,
\label{rr}
\end{equation}
where $C$-s are unknown structure functions of $a$, $b$, $c$ and $d$.
We omit the cubic terms proportional to $\theta^0\theta^1\theta^2$
from the very beginning since they are forbidden by the covariance
condition (\ref{*}).
Note that one has to take into account that on $SL(2)$ the
matrix elements $a,b,c,d$ are not independent:
$\det{T}=ad-bc=1$. Therefore, one of the relations
in eqs. (\ref{rr}) should follow from the three others.
Considering the coordinate patch on
$SL(2)$ where $d\ne 0$ we find:
\begin{equation}
\{\theta^i, a\}=\{\theta^i,
d^{-1}(1+bc)\}=-d^{-1}a\{\theta^i,d\}+
d^{-1}\{\theta^i,b\}c+d^{-1}\{\theta^i,c\}b.
\label{xx}
\end{equation}

The covariance of eq.(\ref{rr})
with respect to the action of $\Delta_R$ (eq.(\ref{*})) leads to the
system of equations on the coefficients $C$:
$$
\Delta
C_{cj}^{i}=C_{cj}^{i}\otimes a+C_{dj}^{i}\otimes c,
$$
\begin{equation}
\Delta C_{dj}^{i}=C_{cj}^{i}\otimes b+C_{dj}^{i}\otimes d,
\label{rr1}
\end{equation}
$$
\Delta C_{bj}^{i}=d^{-1}\left(-aC_{dj}^{i}
+cC_{bj}^{i}+bC_{cj}^{i}\right)\otimes b+
C_{bj}^{i}\otimes d
$$
for any $i,j=0,1,2$. Since the right tensor multipliers in the r.h.s.
of eq (\ref{rr1}) are linear in $a,b,c,d$ we realize that $C$-s are
also linear functions of matrix elements. Then comparing
eqs.(\ref{rr1}) with eqs.(\ref{uu})
one
can easily find the following solutions:
\begin{equation}
C_{cj}^{i}=\alpha_{j}^{i}a+\beta_{j}^{i}c,~~
C_{dj}^{i}=\alpha_{j}^{i}b+\beta_{j}^{i}d,~~
C_{bj}^{i}=-\beta_{j}^{i}b+\gamma_{j}^{i}d,
\label{rr2}
\end{equation}
where $\alpha_{j}^{i},\beta_{j}^{i},\gamma_{j}^{i}$ are arbitrary numerical
coefficients.  Substituting the coefficients (\ref{rr2}) in eq.(\ref{xx}) we
also get the coefficient $C_{aj}^{i}$:
\begin{equation}
C_{aj}^{i}=-\beta_{j}^{i}a+\gamma_{j}^{i}c.
\end{equation}

Having found the form of structure functions we can now use the covariance
of the brackets with respect to the action of $\Delta_L$. For example,
the equation $\{\Delta_L\theta^0,\Delta a\}=\Delta_L\left(
C_{aj}^{0}\theta^j\right)$ gives
\begin{equation}
4abc\otimes a\theta^0+b(1+3bc)\otimes a\theta^2-a^2c\otimes a\theta^1+
b^2d\otimes c\theta^2+abc\otimes c\theta^1+
\label{rr3}
\end{equation}
$$
(1+2bc)a\otimes C_{aj}^{0}\theta^j+b(1+bc)\otimes C_{aj}^{2}\theta^j-
a^2c\otimes C_{aj}^{1}\theta^j+
$$
$$
b(1+2bc)\otimes C_{cj}^{0}\theta^j+b^2d\otimes C_{cj}^{2}\theta^j-
abc\otimes C_{cj}^{1}\theta^j=
$$
$$
\left[\Delta C_{a0}^{0}(1+2bc\otimes I)-\Delta C_{a1}^{0}(2ab\otimes I)+
\Delta C_{a2}^{0}(2cd\otimes I)\right]\left(I\otimes \theta^0\right)+
$$
$$
\left[-\Delta C_{a0}^{0}(ac\otimes I)+\Delta C_{a1}^{0}(a^2\otimes I)-
\Delta C_{a2}^{0}(c^2\otimes I)\right]\left(I\otimes \theta^1\right)+
$$
$$
\left[\Delta C_{a0}^{0}(bd\otimes I)-\Delta C_{a1}^{0}(b^2\otimes I)+
\Delta C_{a2}^{0}(d^2\otimes I)\right]\left(I\otimes \theta^2\right).
$$
The explicit form of $\Delta$-s is taken from eqs. (\ref{rr1}) and
(\ref{rr2}).
Now supposing all the coefficients of linear independent structures in
this equation to be
equal to zero we obtain the set of conditions on the unknown
numerical
coefficients $\alpha_{j}^{i},\beta_{j}^{i}$ and $\gamma_{j}^{i}$.
The surprising fact is that the coefficients are defined from this single
equation up to one continuous parameter, say, $\eta$ (see Appendix).

It remains to check
that this solution satisfies the other eight equations. This tedious but
pure algebraic task can be done and we arrive to
the one-parameter
system of candidates for the brackets.
Note, that the determinant $\det{T}=ad-bc$ is the
central element of the brackets in question as it was desired.

Next we define the values of this parameter $\eta$
from the part of the super Jacobi identities
that involve one $\theta$-variable
and two functions on $SL(2)$:
\begin{equation}
\{\{\theta^i,f\},h\}+\{\{h,\theta^i\},f\}+\{\{f,h\},\theta^i\}=0.
\label{tfh}
\end{equation}
The result is the following: $\eta=0$ or $\eta=-2$.
\begin{lemma}
There exist two and only two graded
Poisson-Lie structures of the
first order on $SL(2)$. They are given by:\\

{\bf I-type}
\begin{equation}
\begin{array}{lll}
\{\theta^0,a\}=-2 c\theta^1,
& \{\theta^1,a\}=a\theta^1, &
\{\theta^2,a\}=4 c\theta^0 -a\theta^2,\\
\{\theta^0,b\}=-2 d\theta^1, &
\{\theta^1,b\}=b\theta^1, &
\{\theta^2,b\}=4 d\theta^0 -b\theta^2, \\
\{\theta^0,c\}=0, &
\{\theta^1,c\}=-c\theta^1, &
\{\theta^2,c\}=c\theta^2, \\
\{\theta^0,d\}=0, &
\{\theta^1,d\}=-d\theta^1, &
\{\theta^2,d\}=d\theta^2,
\end{array}
\label{qqr1}
\end{equation}

{\bf II-type}
\begin{equation}
\begin{array}{lll}
\{\theta^0,a\}=0,
& \{\theta^1,a\}=-a\theta^1, &
\{\theta^2,a\}=a\theta^2,\\
\{\theta^0,b\}=0, &
\{\theta^1,b\}=-b\theta^1, &
\{\theta^2,b\}=b\theta^2, \\
\{\theta^0,c\}=-2a\theta^2, &
\{\theta^1,c\}=4a\theta^0 +c\theta^1, &
\{\theta^2,c\}=-c\theta^2, \\
\{\theta^0,d\}=-2b\theta^2, &
\{\theta^1,d\}=4b\theta^0+d\theta^1, &
\{\theta^2,d\}=-d\theta^2,
\end{array}
\label{qqr2}
\end{equation}
\end{lemma}

The brackets (\ref{qqr1}) and (\ref{qqr2})
have another interesting property, namely,
{\em they are also invariant with respect to the  comultiplication}
$\Delta_G=\Delta_L+\Delta_R$.

\subsection{$\{\theta,\theta\}$ brackets}
Now we are in a position to determine the bracket
containing two $\theta$-s. Having in mind the grading requirement
(\ref{ya2}) one can
write for this bracket a following representation:
\begin{equation}
\{\theta^i,\theta^j\}=C_{mn}^{ij}\theta^m\theta^n,
\label{gg}
\end{equation}
where the coefficients $C_{mn}^{ij}\in \cal A$ are unknown functions
that can be defined by requiring the bracket (\ref{gg}) to obey the
bicovariance conditions (\ref{**}).
Equation for $\Delta_R$ can be written as
$$
\{\theta^i\otimes I,\theta^j\otimes I\}=\{\theta^i,\theta^j\}\otimes I=
\Delta C_{mn}^{ij}(\theta^m\theta^n\otimes I)
$$
and it is obviously satisfied if all the coefficients $C_{mn}^{ij}$ are
pure numbers.

To write down the equations for coefficients $C$ that follows from the
equation for $\Delta_L$  we start with
the bracket $\{\theta^0,\theta^1\}$.

Applying $\Delta_L$ to it's both sides
\begin{equation}
\Delta_L\{\theta^0,\theta^1\}=C_{01}^{01}\Delta_L(\theta^0\theta^1)+
C_{02}^{01}\Delta_L(\theta^0\theta^2)+
C_{12}^{01}\Delta_L(\theta^1\theta^2),
\end{equation}
and using eq.(\ref{**}) we arrive to
$$
-2ab(1+2bc)\otimes \{\theta^0,\theta^0\}-a^3c\otimes\{\theta^1,\theta^1\}-
b^3d\otimes \{\theta^2,\theta^2\}+
$$
$$
8a^2bc\otimes \theta^0\theta^1+4b^2(1+2bc)\otimes \theta^0\theta^1-
2ab(1+4bc)\otimes \theta^0\theta^1+
$$
$$
a^2(1+4bc)\otimes \{\theta^0,\theta^1\}-b^2(3+4bc)
\otimes \{\theta^0,\theta^2\}+ab(1+2bc)\otimes \{\theta^1,\theta^2\}=
$$
\begin{equation}
C_{01}^{01}(a^2\otimes \theta^0\theta^1-ab\otimes
\theta^1\theta^2+b^2\otimes \theta^0\theta^2)+
\label{yr}
\end{equation}
$$
C_{02}^{01}(c^2\otimes \theta^0\theta^1-cd\otimes
\theta^1\theta^2+d^2\otimes \theta^0\theta^2)+
$$
$$
C_{12}^{01}(-2ac\otimes \theta^0\theta^1+(1+2bc)\otimes
\theta^1\theta^2-2bd\otimes \theta^0\theta^2).
$$

Equating the terms containing the linear independent forms $\theta^i\theta^j$
in the right multiplies of tensor product we obtain the system of equations
for coefficients $C$. The solutions of this system are
collected in the \mbox{Table 1:} \\[5mm]
$$
\begin{array}{llllll}
C_{01}^{01}=-2 & C_{02}^{01}=0 & C_{12}^{01}=0 & C_{01}^{00}=\alpha &
C_{02}^{00}=\beta & C_{12}^{00}=\gamma \\
C_{01}^{02}=0 & C_{02}^{02}=2 & C_{12}^{02}=0 &
C_{01}^{11}=0 & C_{02}^{11}=0 & C_{12}^{11}=0 \\
C_{01}^{12}=2\alpha & C_{02}^{12}=2\beta & C_{12}^{12}=4+2\gamma &
C_{01}^{22}=0 & C_{02}^{22}=0 & C_{12}^{22}=0
\end{array}
$$
\begin{center}
Table 1.
\end{center}
Here $\alpha,\beta,\gamma$ are coefficients that are not fixed from the
system. To find $\alpha,\beta,\gamma$ let us consider, for instance, the
l.h.s. of the following Jacobi identity:
\begin{equation}
\{\{\theta^0,\theta^0\},a\}+\{\{a,\theta^0\},\theta^0\}-
\{\{\theta^0,a\},\theta^0\}=0.
\label{ss}
\end{equation}
By using $\{\theta^0,\theta^0\}=\alpha \theta^0\theta^1+\beta \theta^0
\theta^2+\gamma \theta^1\theta^2$ eq.(\ref{ss}) is reduced to the form
$$
(-\alpha(1+\eta)a+2\eta \gamma + 4\eta c)\theta^0\theta^1)+
\beta\eta c\theta^1\theta^2+\beta(1+\eta)a\theta^0\theta^2=0.
$$
Thus, we find $\alpha=0,\beta=0$ for both $\eta=0$ and $\eta=-2$.
Moreover, $\eta(4+2\gamma)=0$, {\em i.e.} $\gamma=-2$ for $\eta=-2$.
Let us show that when $\eta=0$ we still have $\gamma=-2$.
Clearly, taking the equation
$$
\{\{\theta^0,\theta^0\},c\}+\{\{c,\theta^0\},\theta^0\}-
\{\{\theta^0,c\},\theta^0\}=0
$$
when $\eta=0$ and $\alpha=\beta=0$ we obtain:
$(4\gamma+8)a\theta^0\theta^2=0$, {\em i.e.} $\gamma=-2$.

This means that using the bicovariance
condition for the bracket $\{\theta^0,\theta^1\}$ we determine all
the brackets $\{\theta^i,\theta^j\}$ and, moreover, two systems of brackets
(\ref{qqr1}) and (\ref{qqr2}) with $\eta=0$ and $\eta=-2$,
respectively define
only one candidate for the bracket $\{\theta^i,\theta^j\}$. Thus, we
arrive to the Lemma 2.
\begin{lemma}
There is a unique system of brackets of the second order invariant
with respect to the both $\Delta_L$  and $\Delta_R$ and it is given by
\begin{equation}
\begin{array}{lll}
\{\theta^0,\theta^0\}=-2\theta^1\theta^2,&
\{\theta^0,\theta^1\}=-2\theta^0\theta^1,&
\{\theta^0,\theta^2\}=2\theta^0\theta^2, \\
\{\theta^1,\theta^1\}=0,&
\{\theta^1,\theta^2\}=0,&
\{\theta^2,\theta^2\}=0.
\end{array}
\label{dg}
\end{equation}
\end{lemma}

{\em Proof.}
What we have to do is to show that the system of brackets given by
eqs.(\ref{sk}), (\ref{qqr1}) or  (\ref{qqr2}) and (\ref{dg})
satisfies the super Jacobi identity and the
bicovariance condition (\ref{**}).
This is a matter of direct calculations.

 From Lemma 1 and Lemma 2 follows the
\begin {theor}
There exist two and only two graded  Poisson-Lie structures on
the external algebra of  $SL(2)$ prolonging the Sklyanin
brackets (\ref {sk}). They are given by (\ref {qqr1}) or (\ref {qqr2}) and
(\ref {dg})
\end{theor}

\subsection{Differential Poisson structures}
In the theory of Lie groups the operator ${\bf d}$ of exterior derivative plays
an important role.
There is a natural differential operator $\bf d$ on $\cal M$,
${\bf d}:{\cal M}\to{\cal M}$. ${\bf d}$ is defined
on generators of ${\cal M}$ by:
\begin{equation}
{\bf d}T=\theta T~~~{\bf d}\theta=\theta\theta ,
\end{equation}
($\theta$ is the matrix (\ref{theta}) written in terms of
$\theta^i$-generators) and extended to the hole algebra $\cal M$
by using ${\bf d}^2 =0$ and the Leibniz rule.
In principle, graded Poisson structures are not
connected in any way with the operator ${\bf d}$. However, one can rise a
question if there exist such Poisson structures with respect to which ${\bf d}$
is a differentiation.
\begin {defin}
A graded Poisson structure on $\cal M$
is called a differential one if the operator
${\bf d}$ satisfies the Leibniz-like rule:
\begin {equation}
{\bf d}\{ f,h\}=\{ {\bf d}f,h\} +(-1)^{\deg f}\{ f,{\bf d}h\}
\label {LEB}
\end{equation}
\end{defin}

Now we are going to see whether the Poisson-Lie structures on $SL(2)$ given
by (\ref{sk}), (\ref{qqr1}) or (\ref{qqr2}) and (\ref{dg}) are
differential ones. Let us consider for example the bracket $\{ a,b\}
=-ab$. Then \begin{equation} \{ {\bf d}a,b\}+\{ a,{\bf d}b\}=-2ab\theta^0
-(3+2bc)\theta^1 \label {d1} \end{equation} and \begin{equation} {\bf d}\{
a,b\}=-{\bf d}(ab)= -2ab\theta^0 -(1+2bc)\theta^1 .
\label {d2}
\end{equation}
This shows
that our Poisson-Lie structures on $\cal M$ are not the differential ones.
Hence, we arrive to the
\begin{theor}
Differential Poisson-Lie structures on the external algebra of
$SL(2)$ prolonging the Sklyanin brackets do not exist.
\end{theor}

\section{Quantization}
In this section we show that the Poisson-Lie structures on the external
algebra of $SL(2)$ given by eqs.(\ref{sk}), (\ref{qqr1}) or (\ref{qqr2}) and
(\ref{dg}) can be quantized. One defines the quantization of a commutative
graded Poisson-Lie algebra $\cal M$  as a constructing of a
non-commutative algebra ${\cal M}_q$ ($q=\exp h$)
supplied with comultiplications
$\Delta (h), \Delta_{L,R}(h)$ that obeys the following conditions:

1) ${\cal M}_q$ is a free module over
the ring $k[[h]]$, where $h$ is a parameter of quantization,

2) as a
coalgebra ${\cal M}_q/h{\cal M}_q$ is isomorphic to $\cal M$,

3) one can define on ${\cal M}$ the graded Poisson bracket:
\begin{equation}
\{ x, y \}=\lim_{h\rightarrow 0}\left(
\frac{1}{h}[\hat{x},\hat{y}]\right),~~x,y\in {\cal M},~~\hat{x},
\hat{y}\in {\cal M}_q
\label{ren}
\end{equation}

which should coincide with the original bracket on
$\cal M$ \cite{Dr,Fad}.

First of all, note that the quantization of (\ref{sk}) gives the defining
relations describing the function algebra on the quantum group $GL_q (2)$:
\begin{equation}
\begin{array}{lll}
ab=qba & ac=qca & ad-da=\mu bc \\
bc=cb  & cd=qdc & bd=qdb
\end{array}
\label{yy}
\end{equation}
where $\mu =q-1/q$.
The further factorization of $Fun (GL_q (2))$ modulo the quantum
determinant $\det_q T=ad-qbc$ defines the function algebra ${\cal A}_q =Fun
(SL_q (2))$ on the quantum group $SL_q (2)$. The quantization leaves the
coalgebra structure intact.
The quantization of the relations (\ref{qqr1}) and (\ref{qqr2})
and (\ref{dg}) is given by the following theorem.
\begin{theor} Noncommutative algebras generated by the symbols
$\theta^i,~i=0,1,2$ and the generators of the quantum group
$SL_q(2)$ modulo \\
I-type relations:
\begin{equation}
\begin{array}{lll} \theta^0a=a\theta^0+q^2\mu
c\theta^1,& \theta^1a=\frac{1}{q}a\theta^1,& \theta^2a=qa\theta^2-\mu\lambda
c\theta^0, \\ \theta^0b=b\theta^0+q^2\mu d\theta^1,&
\theta^1b=\frac{1}{q}b\theta^1,&
\theta^2b=qb\theta^2-\mu\lambda d\theta^0, \\
\theta^0c=c\theta^0, &
\theta^1c=qc\theta^1, &
\theta^2c=\frac{1}{q}c\theta^2, \\
\theta^0d=d\theta^0, &
\theta^1d=qd\theta^1, &
\theta^2d=\frac{1}{q}d\theta^2
\end{array}
\label{sss}
\end{equation}
or II-type relations
\begin{equation}
\begin{array}{lll}
\theta^0a=a\theta^0,&
\theta^1a=qa\theta^1, &
\theta^2a=\frac{1}{q}a\theta^2, \\
\theta^0b=b\theta^0,&
\theta^1b=qb\theta^1,&
\theta^2b=\frac{1}{q}b\theta^2, \\
\theta^0c=c\theta^0+\mu a\theta^2, &
\theta^1c=\frac{1}{q}c\theta^1-\frac{\mu\lambda}{q^2}a\theta^0, &
\theta^2c=qc\theta^2, \\
\theta^0d=d\theta^0+\mu \theta^2, &
\theta^1d=\frac{1}{q}d\theta^1-\frac{\mu\lambda}{q^2}b\theta^0, &
\theta^2d=qd\theta^2.
\end{array}
\label{sac}
\end{equation}
with $$\mu=q-1/q,~~\lambda =q+1/q,$$
and the relations
\begin{equation}
(\theta^0)^2=\frac{q^2\mu}{\lambda}\theta^1\theta^2,~~
(\theta^1)^2=(\theta^2)^2=0,~~\theta^1\theta^2=-\theta^2\theta^1,
\label{ww}
\end{equation}
$$
\theta^1\theta^0=-\frac{1}{q^2}\theta^0\theta^1,~~~
\theta^2\theta^0=-q^2\theta^0\theta^2.
$$
represent the quantization of the graded Poisson-Lie algebra
(\ref{qqr1}) or (\ref{qqr2}) and (\ref {dg}), respectively.
The quantum determinant
is the central element of the both algebras introduced
above.
\end{theor}

{\em Remark~~} Since the quantization procedure implies that the
comultiplications survive under quantization the Theorem 3 assumes the
existence of the quantum version of left and right comultiplications $\Delta
_{L,R}(q)$.

{\em Proof} At first, one can easily show that in the quasiclassical limit
($h\to 0$) the relations (\ref{sss}) and (\ref{sac}) produce the
Poisson structures of the first and the second types, respectively. Next, one
needs to check the consistency of the algebras introduced in the
theorem with the defining relations of ${\cal A}_q.$
For this purpose it is enough to
consider the ordering of the cubic monomials. Let us, for example, bring
the monomial $ab\theta^0$ into the form $\theta^0ab$ by two different ways:
$$
ab\theta^0=a(\theta^0b-q^2\mu  d\theta^1)=(\theta^0a-q^2\mu
c\theta^1)b-q^2\mu ad \theta^1=\theta^0ab-q\mu\left(q ad+
bc\right)\theta^1,
$$
$$ ab\theta^0=qb(\theta^0a-q^2\mu c\theta^1)=
q(\theta^0b-q^2\mu d\theta^1)a-q^3\mu bc\theta^1=
\theta^0ab-q\mu\left(q da+q^3bc\right)\theta^1.$$
Due to eqs.(\ref{yy})
the results of passing by two different ways are the same. Analogously,
one can show the consistency of the defining relations (\ref{sss}) and
(\ref{sac}) by considering all the other cubic monomials.

Covariance of (\ref{qqr1}), (\ref{qqr2}) and (\ref{dg}) with respect to
$\Delta_L$ and $\Delta_R$ gives rise to the existence of comultiplications
$\Delta_{L,R} (q)$ being
homomorphisms of the algebra described in the Theorem 3.
On $\theta$ generators the action of $\Delta _L(q)$ reads
\begin{equation}
\Delta _L(q)\theta^0=(ad+\frac{1}{q}bc)\otimes\theta^0-qac\otimes \theta^1
+bd\otimes\theta^2,
\label{df12}
\end{equation}
\begin{equation}
\Delta _L(q)\theta^1=-\frac{\lambda}{q}ba\otimes\theta^0
+a^2\otimes\theta^1 -\frac{1}{q}b^2\otimes\theta^2,
\label{df13}
\end{equation}
\begin{equation}
\Delta _L(q)\theta^2=\lambda dc\otimes\theta^0-qc^2\otimes
\theta^1+d^2\otimes \theta^2.
\label{df14}
\end{equation}
and
\begin {equation}
\Delta _R (q)\theta^i=\theta^i \otimes I
\label {u}
\end   {equation}
Such a coproduct is nothing but the adjoint coaction of ${\cal A}_h$ on
$\theta^i$ generators, representing the quantum analog of components of the
right-invariant Maurer-Cartan form. To make it clear let us introduce a
matrix $\theta$
\begin {equation}
\theta=\parallel \theta_{k}^{~m}\parallel=\left(\begin{array}{rr}\theta^0
&\theta^1\\ \theta^2 & -\frac{1}{q^2}\theta^0\end{array}\right) \label {uuq}
\end   {equation}
with
 the quantum trace $\tr{\theta}_q =\frac{1}{q}\theta_{1}^{~1} +q
\theta_{2}^{~2}=0$. Then one can write
\begin {equation}
\Delta_R(q)\theta^i=\theta^i\otimes I,~~~~
\Delta_L(q)\theta_{i}^{~j}=t_{i}^{~k}S(t_{m}^{~j})\otimes \theta_{k}^{~m}
\label {uuu}
\end   {equation}
Now one can show that $\Delta_{L,R}(q)$ are homomorphisms.
At last, using the defining relations
(\ref{sss}) one can check that the quantum determinant $\det_qT$ is the
central element of the algebra under consideration.

\section   {Concluding Remarks and Discussion}
A few remarks are now in order.

$\bullet$
To make the quantization of the brackets
(\ref{qqr1}) and (\ref{dg}) more transparent it is suitable
to rewrite them using the corresponding classical $r$-matrix.  Introducing
the standard tensor notations we have for the brackets of the first order
eq.(3.15) from \cite{AM} with $\alpha=0=\beta$. In the case of $GL(N)$
group these brackets were forbidden by the total Jacobi identity. For the
brackets (\ref{dg})  one can write:
\begin{equation}
\{\theta _1, \theta _2\} =  -(\theta _1\theta _1+ \theta
_2\theta_2)+ r^{12}_{+}\theta _1\theta _2+ \theta _1\theta_2r^{12}_{+}-
\theta _1r^{12}_{+}\theta _2- \theta_2r^{12}_{-}\theta _1,
\label {t13}
\end{equation}
where  $r_{+}$ is the classical $r$-matrix for $sl(2)$ (see (\ref{rrr}))
and $r_{-}=-r_{+}^{t}$.

Let us note that the corresponding quantum algebra (\ref {ww}) was found
in \cite {IS} in the $R$-matrix form
\begin{equation}
R_{12}\theta_1R_{21}\theta_2+
\theta_2R_{12}\theta_1R_{12}^{-1}=
k_q(R_{12}\theta_1^2R_{21}^2+\theta_2^2),
\label{t14}
\end{equation}
where $k_q=\frac{\mu q^2}{q^3+1/q}$ and the quantum $R$-matrix for
$SL_q (2)$ is given in Appendix B. It is clear that the brackets
(\ref{t13}) are the quasiclassical limit of the relations (\ref{t14}).
It seems natural that taken in the $r$-matrix form the graded Poisson-Lie
structures can be generalized to $SL(N)$.

$\bullet\bullet$ Let us consider the
$*$-involution on $SL_q(2)$ defined by \begin{equation}
a^{*}=d,~~d^{*}=a,~~b^{*}=-qc,~~c^{*}=-\frac{1}{q}b.
\label{gmm}
\end{equation}
In addition to (\ref{gmm}) we can introduce the operation $*$ on
$\theta$-s:
\begin{equation}
(\theta^0)^{*}=-\theta^0,~~(\theta^1)^{*}=-\theta^2,~~
(\theta^2)^{*}=-\theta^1.
\label{gm}
\end{equation}
One can easily check that the relations (\ref{ww}) are invariant
with respect to the operation $*$. However, $*$ operation does not
leave both (\ref{sss}) and (\ref{sac}) invariant but transforms the brackets
(\ref{sss}) into (\ref{sac}) and vice versa.

$\bullet\bullet\bullet $
Let us briefly discuss the issue of existence of an operator ${\bf d}_h$ of
exterior derivative for the noncommutative associative algebras described
in the Theorem 3. First of all we would like to stress that this question
has essentially a classical origin.
Obviously, an operator ${\bf d}_h$ should
be introduced in agreement with the defining relations (\ref{yy}) of
${\cal A}_h$ that in the standard tensor notations  read \cite{Fad}:
\begin{equation}
RT_1T_2=T_2T_1R,
\label{gggg}
\end{equation}
where $R$ is the quantum $R$-matrix for $SL_q(2)$ (see (\ref{RRR})).
This means that
\begin{equation}
R{\bf d}_hT_1T_2+RT_1{\bf d}_hT_2={\bf d}_hT_2T_1R+T_2{\bf d}_hT_1R.
\label{gg2}
\end{equation}
Now let us suppose that $R$-matrix and ${\bf d}_h$ operator are
quasiclassical, {\em i.e.}
\begin{equation} R=1+hr+O(h^2),~~~{\bf d}_h={\bf
d}+h{\bf d}^1+O(h^2),
\label{gge}
\end{equation}
where ${\bf d}$ is the ordinary operator of exterior derivative.
Then by using eq.(\ref{ren})
the quasiclassical version of (\ref{gg}) takes the form
$$
\{{\bf d}T_1,T_2\}+\{T_1,{\bf d}T_2\}=[r_{+}^{12}, {\bf d}(T_1T_2)]=
{\bf d}\left([r_{+}^{12}, T_1T_2]\right),
$$
where $\{T_1,T_2\}=[r_{+}^{12}, T_1T_2]$ is the Sklyanin brackets
(\ref{sk}) expressed via the classical $r_{+}^{12}$ matrix for $sl(2)$
(see (\ref{rrr})).
Thus, we conjecture that {\it any noncommutative algebra of quantum
differential forms admits an exterior derivative if the corresponding
graded Poisson-Lie structure is the
differential one}. As we have seen all
graded Poisson-Lie structures on $SL(2)$
are not differential ones and, therefore, the corresponding algebras of
quantum forms do not admit the exterior derivative.

This situation is quite opposite to what we have for $GL(N)$.
In \cite{AM} it was shown that for all graded Poisson-Lie structures
on $GL(N)$ the ordinary {\bf d} operator has the description
in inner terms, namely
\begin {equation} {\bf d}=\{ \kappa\tr{\theta},\ldots\},
\label {dst}
\end {equation}
where $\kappa$ is a coefficient depending on the choice of a Poisson-Lie
structure.  Then the Leibniz rule eq.(\ref{LEB}) reduces to the graded
Jacobi identity \begin {equation} \{\tr{\theta},\{
f,h\}\}=\{\{\tr{\theta},f\},h\} +(-1)^{\deg f}\{ f,\{\tr{\theta},h\}\}
\end{equation}
and, therefore, all graded Poisson-Lie structures on $GL(N)$ appear to be
the differential ones. Moreover, the
possibility of expressing ${\bf d}$ as in (\ref{dst}) allows one to
define its quantum counterpart via the graded commutator $[,]$ as
$$
{\bf d}_h=\left[\frac{1}{\mu}\tr\mbox{_q}{\theta},\ldots \right].
$$

In general, according to
the Connes definition of differentiation of a noncommutative
graded algebra the operator ${\bf d}$ is expressed as
$$
{\bf d}=[\xi ,\ldots ]
$$
where $\xi$ is some special element of degree one such that $[\xi,\xi]=0$.
For a given noncommutative graded
algebra an appropriate element $\xi$ may or may not exist. In the
last case one can try to extend the algebra in such a way as to
include $\xi$ and, therefore to make the algebra a differential one.
Just this situation takes place in the standard approach to  differential
calculi on quantum special groups \cite{Wor},\cite{Car},\cite{WZ}.
In this case the graded algebras of quantum right(left)-invariant forms
can be supplied with the exterior differentiation but the price we have
to pay for doing this consists in extending these algebras by an extra
element $\xi$ which has no natural classical counterpart in the
corresponding classical group. Note, that many of the constructions of
bicovariant differential calculi of such a type can be obtained by the
reduction of bicovariant differential calculi on $GL(N)$
\cite{Car},\cite{WZ}.

$\bullet\bullet\bullet\bullet $
Let us compare now the graded Poisson-Lie structures on $GL(2)$
with the Poisson-Lie structures on $SL(2)$ in more detail.
Among the graded $GL(2)$ Poisson-Lie brackets of the second order
there is only one natural candidate for $SL(2)$ reduction.
Its choice is dictated
by the compatibility with the condition
$\tr{\theta}=0$. This selected Poisson-Lie bracket is given
by the formula (\ref{t13}) where $\theta$ has
four independent entries.
The corresponding quantum algebra
is compatible with the condition $\tr\mbox{_q}{\theta}=0$
and also can be reduced to $SL_q (2)$. It is not surprising that
it is given just by eq.(\ref{t14}).

The situation drastically changes when we search for a graded Poisson-Lie
structure of the first order on $GL(N)$ that can admit
the $SL(N)$-reduction.
It turns out \cite{IS} that
the reduction of any graded Poisson-Lie structure on $GL(N)$
involves inevitably the additional one-form
generator $\xi=\tr{\theta}$. Thus,
the reduction yields the classical limit
of the standard bicovariant
calculi on quantum special groups \cite{Wor},\cite{Car}.
Note, that the relations (\ref{sss}) and (\ref{sac}) change the
transformation property of (\ref{dg}) as compared with the
$GL_q(2)$ case. Namely, despite the fact that
\begin{equation}
\Delta_G (q)=\Delta_L(q)+\Delta_R(q)
\label{qqq}
\end{equation}
is the homomorphism of the algebras (\ref{sss}) and (\ref{sac}) the
relations (\ref{dg}) in opposite to the $GL(N)$ case do not
respect the action of (\ref{qqq}).

Finally, one can address a question whether it is possible to find for the
algebras (\ref{sss}) and (\ref{sac}) an $R$-matrix formulation. Note,
that the existence of such a formulation can make the generalization
of our $SL(2)$ constructions to $SL(N)$ case straightforward.

$$~$$
{\bf ACKNOWLEDGMENT}
$$~$$
The authors are grateful to I.Volovich for interesting
discussions.
This work is supported in part by RFFR under grant N93-011-147.
$$~$$
\newpage

{\large \bf APPENDIX}
\appendix
\section{One-parameter family of candidates for graded Poisson-Lie
structure on $SL(2)$}
\setcounter{equation}{0}
\begin{equation}
\begin{array}{ll}
\{\theta^0,a\}=\eta c\theta^1, &
\{\theta^1,a\}=-(1+\eta) a\theta^1, \\
\{\theta^0,b\}=\eta d\theta^1, &
\{\theta^1,b\}=-(1+\eta)b\theta^1, \\
\{\theta^0,c\}=-(2+\eta)a\theta^2, &
\{\theta^1,c\}=(4+2\eta)a\theta^0+(1+\eta)c\theta^1,\\
\{\theta^0,d\}=-(2+\eta)b\theta^2, &
\{\theta^1,d\}=(4+2\eta)b\theta^0+(1+\eta)d\theta^1,
\end{array}
\label{qqr}
\end{equation}
$$
\begin{array}{l}
\{\theta^2,a\}=-2\eta c\theta^0+(1+\eta)a\theta^2, \\
\{\theta^2,b\}=-2\eta d\theta^0+(1+\eta)b\theta^2,  \\
\{\theta^2,c\}=-(1+\eta)c\theta^2, \\
\{\theta^2,d\}=-(1+\eta)d\theta^2.
\end{array}
$$
\section{The quantum $R$-matrix for $SL_q(2)$}
\setcounter{equation}{0}
\begin{equation}
R=\left(\begin{array}{llll}q & 0 & 0 & 0 \\
0 & 1 & 0 & 0 \\
0 & \mu & 1 & 0 \\
0 & 0 & 0 & q \\
\end{array}\right)
\label{RRR}
\end{equation}
\section{The classical $r_{+}^{12}$ matrix for $sl(2)$}
\setcounter{equation}{0}
\begin{equation}
r_{+}=\left(\begin{array}{llll}1 & 0 & 0 & 0 \\
0 & 0 & 0 & 0 \\
0 & 2 & 0 & 0 \\
0 & 0 & 0 & 1 \\
\end{array}\right)
\label{rrr}
\end{equation}

\end{document}